\documentclass[12pt,english,aps,prb,preprint,showpacs]{revtex4}
\usepackage{amsmath}
\usepackage{graphicx}
\providecommand{\tabularnewline}{\\}
\usepackage{bm}
\usepackage{babel}
\makeatother
\begin{document}

\title{Simple models for 2D tunable colloidal crystals in rotating AC Electric
fields}

\author{Nils Elsner}

\affiliation{School of Chemistry, University of Bristol, Bristol BS8 1TS, UK}

\altaffiliation{VDL, Fuchstan str. 61, Frankfurt am Main, 60489, Germany.}

\author{David R.E. Snoswell}

\affiliation{Department of Physics, Cavendish Laboratory, University of Cambridge,
J.J. Thomson Avenue, Cambridge CB3 0HE, UK}

\author{C. Patrick Royall}

\email{paddy.royall@bristol.ac.uk}

\affiliation{School of Chemistry, University of Bristol, Bristol BS8 1TS, UK}

\altaffiliation{Institute of Industrial Science, University of Tokyo, 4-6-1 Komaba, Meguro-ku, Tokyo 153-8505, Japan}

\author{Brian Vincent}

\affiliation{School of Chemistry, University of Bristol, Bristol BS8 1TS, UK}

\date{\today}

\begin{abstract}
We compare the behaviour of a new 2D aqueous colloidal model system
with a simple numerical treatment. To first order the attractive interaction
between the colloids induced by an in-plane rotating AC electric field
is dipolar, while the charge stablisation leads to a shorter ranged,
Yukawa-like repulsion. In the crystal-like `rafts' formed at sufficient
field strengths, we find quantitative agreement between experiment
and Monte Carlo simulation, except in the case of strongly interacting
systems, where the well depth of the effective potential exceeds 250
times the thermal energy. The `lattice constant' of the crystal-like
raft is located approximately at the minimum of the effective potential,
resulting from the sum of the Yukawa and dipolar interactions. The
experimental system has display applications, owing to the possibility
of tuning the lattice spacing with the external electric field. Limitations
in the applied field strength and relative range of the electrostatic
interactions of the particles results in a reduction of tunable lattice
spacing for small and large particles respectively. The optimal particle
size for maximising the lattice spacing tunability was found to be
around 1000 nm. 
\end{abstract}


\maketitle

\section{Introduction}

\label{introduction}

Colloidal suspensions present the possibility to develop novel materials
via self-assembly. Of particular interest are colloidal crystals,
whose optical properties can generate iridescent colours, and provide
a means by which photonic crystals may be produced\citet{joannopoulos1997},
while further applications range from lasers \citet{colombelli2003}
to display devices \citet{aresnault2003}, with recent advances demonstrating
tunable colours through control of lattice spacing with an external
field \citet{arsenault2007}. Further to the practical importance
of colloidal crystals, their well-defined thermodynamic temperature
allows colloidal dispersions to be viewed as mesoscopic `model atoms'
\citet{pusey}.

Recently, the ability to tune the colloid-colloid interactions has
led to the observation of a wide variety of structures \cite{jones1995,yethiraj2003,manoharan2003,martin2003PRL,leunissen2005,snoswell2006}.
Of particular interest here, to first order AC electric fields can
induce dipolar interactions between the colloidal particles, leading
to anisotropic interparticle potentials and exotic crystal structures,
some of which are not observed in atomic and molecular systems \citet{yethiraj2003},
while external control of the colloid-colloid interactions allows
direct observation of phase transitions \cite{yethiraj2004}. Furthermore,
direct microscopic observation at the single-particle level allows
an unprecedented level of detail to be accessed \cite{vanblaaderen1995},
opening the possibility of tackling long-standing problems in condensed
matter, such as freezing \cite{gasser2001}.

The introduction of a rotating AC field opens up even more possibilities.
In this case, the dipolar interactions lead to an attraction in the
plane of rotation and to repulsions above and below. Studies with
a rotating magnetic field on granular matter indeed produced disc
like patterns consistent with expectations \cite{halsey1995,martin1999}.
Unlike granular matter, since colloidal dispersions exhibit Brownian
motion, thermodynamic equilibrium structures (ie crystals), may be
obtained \citet{lumsdon2003,snoswell2006}. In previous work Snoswell
\emph{et.al.} \citet{snoswell2006} showed that lattice spacing within
quasi-2D colloidal crystals could be controlled \emph{in-situ}, by
means of coplanar rotating electric field. The interparticle dipolar
interactions in the plane of the electric field may be treated to
first order as a circularly symmetric attraction, due to the time
averaging effect of a rapidly rotating field (1000Hz) on relatively
large particles on the micron lengthscale, where the diffusive timescale
is of the order of seconds \citet{russell}.

In considering the interactions between the particles, the asymmetry
between the colloids (10nm-1$\mu$m) and smaller molecular and ionic
species must be addressed. A number of coarse-graining schemes have
been developed where the smaller components are formally integrated
out \citet{likos2001}. This generates a one-component picture, where
only the effective colloid-colloid interactions are considered, and
the complexity of the description is vastly reduced. The equilibrium
behaviour of the colloids in the original multi-component system may
then be faithfully reproduced by appeal to liquid state theory \citet{hansen}
and computer simulation \citet{frenkel}. Central to the success of
this one-component approach is the use of a suitable effective colloid-colloid
interaction $u(r)$.

In this study, we use a simple numerical treatement in which we can
predict the lattice spacing in the quasi 2D crystal from the electric
field strength. We consider a model system of charged colloids, in
a rotating electric field \citet{snoswell2006}. By exploiting the
knowledge both of the electrostatic repulsions and dipolar attractions,
we present a direct, quantitative comparison of a tunable interaction
and a material property of the crystalline `rafts' formed. We combine
experimental measurements of the crystal lattice constant $d$ as
a function of field strength $E$ with Monte-Carlo simulations according
to a screened Coulomb repulsion plus dipolar attraction where the
only fitting parameter is the Debye screening length.

In the simulations, we use pairwise interactions, in other words we
assume that at the higher densities at which the crystalline rafts
are formed, the system is still accurately described by interactions
calculated for two particles in isolation. We note that deviations
from this assumption of pairwise additivity have been measured both
in the case of strongly charged colloids \citet{brunner2002} and
in the case of \emph{repulsive} dipolar interactions \cite{zahn2003}.
We further compare simulation results with the minimum of the effective
potential, which we take as a measure of the lattice constant of the
crystalline rafts, which we also determine from experimental data.

This paper is organised into six sections. In section \ref{theory}
we present expressions for the effective interactions between the
colloids, summing the attractions and repulsions to provide an effective
one-component description of the system. Section \ref{experimental}
describes our experimental metholodogy. Section \ref{simulation}
outlines the Monte-Carlo simulation technique employed. The comparison
of simulation and experimental results is presented in section \ref{results}
and in section \ref{tunability} we extrapolate our findings to maximise
the tunability of the crystal lattice constant, which may be useful
for applications. We conclude our findings in section \ref{conclusions}.

\section{Theory and Model}

\label{theory}

In the following we will consider a system consisting of two particles
in a surrounding medium. We shall assume that these particles are
charged, leading to a repulsive interaction, and that the rotating
AC electric field induces a dipole moment in the two particles and
thus induces an attractive interaction. To describe this system, we
start from the Derjaguin, Landau, Verwey and Overbeek (DLVO) approach
\cite{verwey1948}, which consists of attractive van der Waals interactions
at short range, and long-ranged repulsive electrostatic interactions.
The van der Waals interactions are very short-ranged, and are neglected,
as electrostatic repulsions inhibit the close approach at which van
der Waals interactions become important. We shall therefore assume
that the only relevant attractions result from the long-ranged dipolar
interactions induced by the rotating electric field.

In the linear Poisson-Boltzmann regime, the electrostatic repulsions
may be expressed as a hard core Yukawa, or screened Coulomb interaction
\cite{verwey1948},

\begin{equation}
\beta u_{yuk}(r)=\begin{cases}
\text{$\infty$} & \text{for $r<\sigma$}\\
\text{$\beta\epsilon_{yuk}\frac{\exp(-\kappa(r-\sigma))}{r/\sigma}$} & \text{for $r\ge\sigma$}\end{cases}\label{eqYuk}\end{equation}

\noindent where $\beta=1/k_{B}T$ where $k_{B}$ is Boltzmann's constant,
$T$ is temperature and $\sigma$ is the colloid diameter. The potential
at contact, $\beta\varepsilon_{yuk}$ is given by

\begin{equation}
\beta\epsilon_{yuk}=\frac{Z^{2}}{(1+\kappa\sigma/2)^{2}}\frac{l_{B}}{\sigma}\label{eqEpsilonYuk}\end{equation}

\noindent where $Z$ is the colloid charge, $l_{B}$ is the Bjerrum
length and $\kappa=\sqrt{4\pi l_{B}\rho_{i}}$ is the inverse Debye
screening length where $\rho_{i}$ is the total number density of
monovalent ions.

Now the regime of linear Poisson-Boltzmann theory in which equation
(\ref{eqYuk}) holds corresponds to relatively weak charging. Although
this is not the case here, a potential of the Yukawa form is recovered
at larger separations, if a smaller, renormalized charge is considered
\citet{alexander1984}. We tabulate measurements of the $\zeta$-potential
of dilute suspensions in table \ref{tableKappaSigma}. These values
suggest that we expect a renormalised charge, in the conditions under
which the colloids form the crystalline rafts, rather high colloid
concentration, hence high counter ion concentration \cite{russell,alexander1984}.
Therefore, noting that the Debye length is much smaller than the colloid
radius, we follow Bocquet \emph{et. al.} \cite{bocquet2002} and take
the following expression for the renormalised charge $Z_{eff}$ :

\begin{equation}
Z_{eff}=\frac{4\sigma}{l_{B}}\frac{(1+\kappa\sigma/2)^{2}}{1+\kappa\sigma}\label{eqZeff}\end{equation}
 which we substitute for $Z$ in equation (\ref{eqEpsilonYuk}). This
expression gives good agreement with measurements of the effective
colloid charge for particles with a comparable $\zeta$-potential
\cite{yamanaka1990}. We recall that many-body effects can lead to
a density-dependence in the effective colloid-colloid interactions
\citet{brunner2002,zahn2003}. However, we shall neglect these effects
in the present work.

The attractive potential between the colloids resulting from the rotating
AC electric field is treated to first order as a dipolar interaction:
\begin{equation}
u_{dip}(r)=\frac{(p\times p)}{2\pi\epsilon\epsilon_{0}r^{3}}\label{eqDip}\end{equation}

\noindent where $p$ is the induced dipole moment of each particle,
$\epsilon$ is the dielectric constant and $\epsilon_{0}$ is the
permittivity of free space. The dipole moment may be calculated from
the strength of the electric field

\begin{equation}
p=\frac{1}{2}\pi\epsilon\epsilon_{0}\sigma^{3}K\underline{E}\label{eqMoment}\end{equation}

\noindent where $K$ is the Clausius-Mosotti factor and takes values
between 1 and -1/2 depending upon the origin of the dipole moment.
We note that in this case of an alternating field, the root-mean-square
of the time dependent field is taken. Substituting $r=\sigma$ leads
us to a potential at contact

\begin{equation}
\epsilon_{dip}(\sigma)=\frac{(p\times p)}{2\pi\epsilon\epsilon_{0}\sigma^{3}}\label{eqEpsilonDip}\end{equation}

\noindent which has a cubic dependence on the colloid diameter, in
the case of all other contributions being unchanged.

In considering the value of the Clausius-Mosotti factor, two regimes
are relevant, corresponding to applied electric fields of high and
low frequency, in which the dipolar interactions result from dielectric
permittivity differences between the particles and the solvent and
differences in conductivity respectively \cite{jones1995}. The crossover
frequency $\omega^{*}$ between these regimes is given by

\begin{equation}
\omega^{*}=\frac{1}{2\pi\epsilon_{0}}\sqrt{-\frac{s_{p}^{2}-2s_{p}s_{m}+2s_{m}^{2}}{\epsilon_{p}^{2}-\epsilon_{p}\epsilon_{m}+2\epsilon_{m}^{2}}}\label{eqCrossover}\end{equation}

\noindent where $s$ are conductivities and the indices $p$ and $m$
refer to the colloidal particles and the medium respectively. Following
\citet{ermolina2005} the conductivity of the particles $s_{p}$ is
taken to be the sum of the bulk conductivity $s_{b}\approx0$ and
the conductivity on the surface $s_{s}$.

\begin{equation}
s_{p}=s_{b}+s_{s}\sigma\label{eqConductivity}\end{equation}

In principle equation (\ref{eqConductivity}) holds for particles
larger than 1 $\mu$m. In the case of much smaller particles, one
should take into account that the contribution of the diffuse layer
to the conductivity of the particles. Noting that the smallest particles
used in the experiments are 757 nm in diameter and that the conductivity
of the de-ionised water is relatively low, neglect the double layer
contribution and thus use equation (\ref{eqConductivity}). Since
the largest particles studied are 2070 nm in diameter and the frequency
is 1 kHz and the bulk ionic strength is $0.01-0.1$ mMol as determined
from conductivity measurements, we work in the low frequency regime
where the contributions from the conductivities dominate \citet{ermolina2005}.
The Clausius-Mosotti factor therefore takes the following form:

\begin{equation}
K_{s}=\frac{s_{p}-s_{m}}{s_{p}-2s_{m}}\label{eqClausius}\end{equation}

\noindent Due to the surface conductivity, $s_{p}$ is much larger
than the conductivity of the medium $s_{b}$. The value of $K_{s}$
is therefore close to one. We now combine the contributions from the
electrostatic repulsions and the dipolar attractions, yielding the
expression.

\begin{equation}
u_{tot}(r)=u_{yuk}(r)+u_{dip}(r)\label{eqU}\end{equation}
 $u_{tot}$ has the form indicated in Fig. \ref{figU}. We see a minimum
in the potential, which one might expect to provide a first approximation
to the lattice constant in the 2D colloidal crystal. In sections \ref{results}
and \ref{tunability} we shall compare this minimum to our simulation
results.

\section{Experimental}

\label{experimental}

Colloidal crystals underwent self-assembly as a result of an applied
electric field. We used anionic, sulphate stabilised polystyrene latex
particles, either synthesised using a standard technique, surfactant
free emulsion polymerisation \citet{goodwin1973} in the case of $\sigma=$757
and 945 nm, or particles produced by the same method but purchased
from Microparticles GmbH for $\sigma=$1390 and 2070 nm.

Particle electrophoretic mobility was measured using a Brookhaven
Zetaplus light scattering instrument. Particle sizes were determined
by scanning electron microscopy, either in-house using a Jeol JSM-6330F,
in the case of $\sigma=$757 and 945 nm or by Microparticles GmbH
for $\sigma=$1390 and 2070 nm are are listed in table \ref{tableKappaSigma}.
Experiments were performed with dilute aqueous suspensions (0.5-1.5
wt\%). Schematics of the experimental set up are shown in Fig. \ref{figSchematic}
All glass surfaces were chemically washed with 0.1M KOH and
washed with copious quantities of MilliQ water. Particles were deionised
by direct contact with ion exchange beads before being made up to
the desired electrolyte concentration with KCl. Very low
salt conditions are required as even moderate (mMol) salt concentrations
lead to electrohydrodynamic pattern formation due to ion flow \cite{isambert1997,lumsdon2004}.
For further details of the experimental set up and procedure the reader
is referred to Snoswell \emph{et. al.} \cite{snoswell2006}.

We consider our experiment as a 2D system. However, occasionally we
noticed some overlap of the crystalline rafts. We do not include these
results in our analysis. In order to treat the system in 2D, one might
expect the gravitational length $l_{g}=k_{B}T/mg$, where $m$ is
the bouyant mass of the particle and $g$ is the acceleration due
to gravity, to be much less than some characteristic length such as
the particle diameter. Due to the fairly small density mismatch between
polystyrene and water, in fact even for the largest particles $\sigma=2070$
nm, $l_{g}\approx0.87\sigma$. However, the dipolar interactions between
the particles are attractive in-plane, but strongly repulsive in the
vertical direction. This promotes the formation of `rafts' and `sheets',
and these large assemblies of many particles have very small gravitational
lengths. Thus we argue that the system behaves in a quasi 2D manner.
The lattice parameter $d$ was taken as the average of typically
ten crystalline `rafts', measured across the `raft', of around ten
lattice spacings. The response to changing the electric field strength
was determined to be less than 100 ms, we waited 5s after changing
the field strength before acquiring data.

\section{Monte-Carlo Simulation}

\label{simulation}

We use standard Monte-Carlo (MC) simulations in the NVT ensemble \cite{frenkel}.
The particles interact via equation (\ref{eqU}) in two dimensions.
To mimic the experiments, we initialise the system in a random configuration
at a relatively low concentration, corresponding to an area fraction
$\phi=0.05-0.3$. We confirmed that different area fractions gave
indistinguishable results, however smaller area fractions often led
to longer equilibration times. The attractive interactions cause the
particles to approach one another, and form crystallites which then
coalesce to form a crystalline `rafts' which contain of order $100$
particles, in a qualitatively similar way to the experimental system.
Each simulation was equilibrated for typically 30000 MC moves per
particle, followed by 3000 production moves per particle. We recall
that a potential of the form $1/r^{3}$ converges in 2D. The lattice
constant for this 2D hexagonal crystal was taken as

\begin{equation}
d=\frac{\int_{0}^{b}rg(r)dr}{\int_{0}^{b}g(r)dr}\label{eqLattice}\end{equation}

\noindent where $g(r)$ is the pair correlation function, and $b$
the minimum between the first and second peaks of $g(r)$.

We found that when we used $N<200$, that typically one crystalline
domain was formed, around the centre of the simulation box. We therefore
do not use periodic boundary conditions, and consider the lattice
spacing of the single crystal formed in the simulation. One parameter
is the number of particles typically present in each crystalline domain.
This is estimated to be around $N=100$$\pm25$ from experimental
data (see Fig. \ref{figPix}). This value is governed by the
overall concentration of the colloidal suspension. We have considered
the effects of varying $N$ as shown in Fig. \ref{figFiniteSize}.
The results of a considerable range of $N$ resulted in a variation
of a around one percent in $d$ for the $\sigma=945$ nm system for
a field $E$=20 $kVm^{-1}$. Each simulation was repeated four times
except $N=576$. We found a slight trend to tighter binding for larger
$N$, however, as shown in Fig. \ref{figFiniteSize}, this effect
is rather small. We henceforth use one simulation per state point
unless otherwise stated. The slight scatter in the simulation data
is perhaps indicative the existence of different 'magic numbers' for
these crystalline rafts \cite{wales}. These may be thought of as
2D hexagonal clusters, whose `magic numbers', ie low energy states,
are expected to include $7$, $19$, $37$... Close to a `magic number'
the binding may be expected to be relatively tighter. A more detailed
exploration of this phenomenon lies beyond the scope of this work.
Overall, the variation in $d$ as a function of $N$ is smaller than
the experimental scatter, nonetheless we take both $N=72$ and $N=144$
values for $d$ when comparing with experimental data. The slightly
larger value for $N=32$ in Fig. \ref{figFiniteSize} is attributed
to the small size of the crystalline `raft', such that the surface
particles which are more widely spaced make a greater contribution.
The domain size is of order 100 particles, thus, for larger $N$,
defects and grain boundaries may lead to a smaller contribution to
the measurement of $d$. This also applies to experimental data.

For fitting, all parameters are known, except the Debye length $\kappa^{-1}$
which we take as a free parameter for each particle size and salt
concentration, although we note that the effective colloid charge
is itself a function of the Debye length, equation (\ref{eqZeff}).
It has previously been shown that comparison of simulation and experimental
data can yield a reasonable \emph{local} measure of the Debye length
\cite{royall2006}. Furthermore, in the region of the sample in which
the electric field is applied, the colloid volume fraction is much
higher, and due to the colloidal counter-ions the ionic strength may
increase, leading to a reduction in the Debye length with respect
to the bulk. However, it is the Debye length in this region which
is relevant for the effective colloid-colloid interactions. The bulk
Debye length that may be determined, for example from conductivity
measurements, may therefore be taken only as an upper bound. System
parameters for different particle sizes are tabulated in table \ref{tableKappaSigma},
and the Debye length is plotted in Fig. \ref{figKappaSigma}(a) and
the contact potential of the Yukawa interaction {[}equation (\ref{eqEpsilonYuk})]
in Fig. \ref{figKappaSigma}(b) are plotted as a function of particle
diameter.

Clearly, our MC simulations may be expected to provide a reasonable
treatment of the crystalline rafts in (quasi) equilibrium, rather
than to describe the formation process. We therefore restrict our
analysis to a simple characterisation of the crystal rafts, and primarily
consider the lattice spacing. We furthermore note both here and in
previous work \cite{snoswell2006} that considerable variation in
shapes of crystalline rafts was observed, but that this had little
impact upon the measurement of the lattice spacing $d$. Likewise,
at the level of this work, we neglect possible local variations in
$d$ due to the proximity of an interface. In any case we note that
for display applications, all lattice spacings contribute to the diffraction.
A few words on equilibration are in order. This applies both
to the experimental system, and to the simulations. Neither system
is strictly in equilibrium, in that case we might expect a rather
regularly-shaped raft such as a hexagon. However, the insensitivity
of the lattice parameter either as a function of $N$ (Fig. \ref{figFiniteSize}),
and the very close agreement between statistically independent simulation
runs lead us to conclude that our approach is sufficient to compare
lattice parameters between experiment and simulation.

\section{Results}

\label{results}

The effect of changing the electric field strength is readily demonstrated
in Fig. \ref{figPix}, which shows optical microscope images of the
region of the sample in which the field is applied. Two field strengths
are shown, 29 kVm$^{-1}$ (a) and 80 kVm$^{-1}$ (b) for $\sigma=945$
nm and we see a correspondingly tighter lattice in the case of the
higher field strength. Similar behaviour was observed in \cite{snoswell2006}.
Figure \ref{figPix}(c) illustrates simulated coordinates for $\varepsilon_{yuk}$$=5.78\times10^{3}$
$k_{B}T$, $\kappa\sigma=28.4$, $\varepsilon_{dip}$$=50.1$ $k_{B}T$,
which corresponds to $E$=29.0 Vm$^{-1}$for the $\sigma=945$ nm
system (see table \ref{tableKappaSigma}).

A more quantitative comparison between experiment and simulation is
shown in Fig. \ref{figG}, which plots the radial distribution function
$g(r)$. We see that the structure appears to be crystalline, with
reasonably long-ranged correlations, although, as is clear from Fig.
\ref{figPix}, the system is too small to test for truly long-ranged
order. We thus note that it is difficult to distinguish the true crystal
from a hexatic phase in this system. However, our motivation is to
model the experimental system, with $N=100\pm25$ and thus we argue
that identification of the hexatic phase lies beyond the scope of
this work.

Having illustrated the general behaviour of the system, we now turn
our attention to the fitting of the experimental data, with the model
described in section II. The experimental data was fitted to the theory
by taking the Debye-length as the fitting parameter.

The main results are shown in Figs. \ref{figSmallExpMC} and \ref{figBigExpMC}
for $\sigma=757$ and 945 nm, and $\sigma=1390$ and 2070 nm systems
respectively. These concern the lattice parameter $d$ as a function
of applied field strength $E$. In general, the simulation is able
to capture the behaviour of the system in a reasonably quantitative
manner. Furthermore, simply plotting the minimum in the potential
$u_{min}$ given by equation (\ref{eqU}) provides a first approximation
to the lattice constant $d$. Figure $\ref{figSmallExpMC}$(a) shows
the results for $\sigma=757$nm particles. Due to the small particle
size, and the inverse cubic dependence of the strength of the dipolar
attractions on the particle diameter, as characterised by $\epsilon_{dip}$
{[}equation (\ref{eqEpsilonDip})] the attraction for a given field
strength is comparitively small, so we find relatively larger lattice
parameters for this system. At low field strengths ($E<20$ $kVm^{-1}$)
we see an increase of the lattice constant for both the experimental
data and the simulation, compared to the minimum in the potential.
Apparently, as the crystal approaches melting, fluctuations become
more pronounced, leading to an increase in $d$, which is not captured
in equation (\ref{eqU}), or perhaps a molten surface layer increases
the apparent value of $d$. We leave this intriguing question for
further investigations. Meanwhile, at high field strengths, we find
some deviation between simulation, experiment and equation (\ref{eqU}).
Apparently, due to the long ranged nature of the interaction, second
nearest neighbours experience an attraction of sufficient strength
that the crystal is compressed by its own cohesion, leading to a smaller
lattice constant.

In the case of the $\sigma=757$ nm particles, we find that for small
field strengths, ($E=10$ $kVm^{-1}$), the system does not form crystallites,
rather it remains as a colloidal liquid. We determined this both by
experimental observation and with simulations. In the latter case,
we identified crystallisation with a splitting in the second peak
of the pair correlation function $g(r)$. The kink around $E\sim20$
$kVm^{-1}$in Fig. \ref{figSmallExpMC} is likely related an artifact
of measuring multiple rafts, so defects and grain boundaries contribute
to the value of $d$. At weak field strengths the rafts tend to be
smaller, thus increasing this effect.

At larger particle sizes, (Fig. \ref{figBigExpMC}), we again see
reasonable agreement between experiment and simulation, albeit with
some deviation between simulation and equation (\ref{eqU}) at higher
field strengths, as the simulation predicts a smaller lattice constant
than that observed in the experiments. However, for the $\sigma=1390$
nm system in particular, in fact equation (\ref{eqU}) provides a
more accurate description of the experimental data. Apparently some
of our assumptions in the simulations, perhaps that of pairwise additivity,
begin to break down at high field strengths. According to equation
(\ref{eqU}), the potential at the minimum of the attractive well
is some 250 $k_{B}T$ for an applied field of 50 kVm$^{-1}$ for $\sigma=1390$
nm. Thus we conclude that for moderate interaction strengths, equation
(\ref{eqU}) provides a reasonable description of the system.

We now consider the dependence of the system upon the colloid diameter
$\sigma$. Firstly, we see from Fig. \ref{figKappaSigma}(a) that
the absolute value of the (fitted) Debye length $\kappa^{-1}$ does
not change significantly for all the four particle sizes studied.
Thus, the reduced inverse Debye length $\kappa\sigma$ is linear in
$\sigma$. The contact potential $\beta\epsilon_{yuk}$ has an approximately
inverse cubic dependence on $\sigma$ for $\kappa\sigma$$\gg1$,
equation (\ref{eqEpsilonYuk}), but also depends upon the (effective)
charge $Z_{eff}$. Plotting $\beta\epsilon_{yuk}$ as a function of
$\sigma$ in Fig. \ref{figKappaSigma}(b) we find an approximate power
law dependence with an exponent of $b=0.91\pm0.01$.

As noted above, the prefactor of the dipolar attraction, $\epsilon_{dip}$
{[}equation (\ref{eqEpsilonDip})], has a $\sigma^{-3}$ dependence.
Although the electrostatic interactions are non-negligible, we nonetheless
expect from equation (\ref{eqEpsilonDip}) that upon decreasing $\sigma$
we find a relatively larger lattice constant for a given field strength,
and that larger field strengths are required to provide sufficient
interactions that the crystalline rafts form. This we indeed find,
as shown in Figs. \ref{figSmallExpMC} and \ref{figBigExpMC}, and
also in the next section.

\section{Optimising the tunability}

\label{tunability}

Now we have noted that this system has optical display applications,
due to the possibility of externally tuning the lattice parameter
$d$. These applications may be most usefully realised when the possibility
to tune the system is maximised, ie when the range of $d$ is maximised.
We are therefore motivated to consider a range of particle sizes,
and calculate the range of tunability of $d$. Now $\sigma$ sets
a lower bound to $d$ while the upper bound is set by the melting
transition which can be determined by Monte-Carlo simulation.

Hitherto, we have used the Debye length $\kappa^{-1}$ as a fitting
parameter. However, as Fig. \ref{figKappaSigma}(a) shows, there is
relatively little change in the absolute value of the Debye length
for the range of colloid diameters investigated. We therefore fix
$\kappa^{-1}=40$ nm and vary the colloid size, and apply the same
methodology as that outlined in sections \ref{theory} and \ref{simulation},
in calculating the response for $\sigma=400-2000$ nm colloids to
an external electric field. While an accurate determination of the
melting transition is complicated by the small system size, melting
is approximately determined by the splitting of the second peak in
the radial distribution function $g(r)$ \cite{truskett1998}. We
decrease the electric field to identify the weakest field at which
the second peak in $g(r)$ exhibits clear splitting {[}Fig. \ref{figTuning}(a),
inset], thus yielding $d(E_{min})$, the lattice spacing just prior
to melting. The melting value of the lattice parameter $d_{m}$ is
taken as

\begin{equation}
d_{m}=d(E_{min})+\frac{d(E_{min}+\delta E)}{2}\label{eqMelting}\end{equation}
 where $\delta E$ is the amount by which the field strength is varied
between simulations, typically $0.5-1.0$ $kVm^{-1}$. The error in
$d_{m}$ is then taken as $d$$_{m}-d(E_{min})$.

In terms of the particle diameter, $d_{m}$ grows considerably at
small sizes, as the relatively larger Debye length leads to a minimum
in the effective potential at relatively larger distances. However,
we recall that our analysis suggests a roughly $\sigma^{3}$ dependence
of $\epsilon_{dip}$ the strength of the dipolar attraction. This
suggests that we might expect a stronger electric field to be required
for relatively small colloids, and indeed we find that a higher field
is required to provide sufficient cohesive energy to hold the colloids
together in the crystal-like `rafts', so at 100 $kVm^{-1}$, a value
we take as a reasonably accessible maximum field strength, the lattice
spacing is still around $d_{100}\approx1.6$$\sigma$ for $\sigma=400$
$nm$, where $d_{100}$ is the lattice spacing corresponding to a
field strength of 100 $kVm^{-1}$. Conversely, at larger particle
sizes, the Debye length is relatively small, so we find, even at melting,
that the lattice constant only approaches around $1.2$$\sigma$.
We note that, at high field strengths, the assumptions of
Section \ref{theory} will ultimately break down. The results presented
in Figs. \ref{figSmallExpMC} and \ref{figBigExpMC} show that $E<100$
$kVm^{-1}$ is reasonable, at least for $\sigma>757$ nm, supporting
our conclusion of a maximum in tuneability around $1.2$$\sigma$.
We show the values of the well depth of equation (\ref{eqU}) at melting
for the different experimental system in table \ref{tableKappaSigma},
which indicates a trend towards deeper wells for larger particles,
i.e. shorter ranges of the repulsive interaction relative to the particle
size.

Larger field strengths may be expected to yield greater tunability
for the small particle sizes. Since the overall minimum lattice spacing
is $d=\sigma$, in principle smaller particles have greater tunability,
however experimental observations reveal intense fluid turbulance
disrupts crystal formation at higher field strengths. This is caused
by large field gradients at the electrode edges that induce strong
dielectrophoretic forces. In addition, higher field strengths can
cause bubble formation by electrolysis and electrochemical degradation
of the electrodes.

The result is thus that intermediate particle sizes have the highest
degree of tuneability, for a given maximum field strength. Here we
have considered 100 $kVm^{-1}$, which leads to a maximum tuneability
of $d_{m}/d{}_{100}$ in the range 1000 nm $<\sigma<$1500 nm.

\section{Conclusions}

\label{conclusions}

2D Colloidal crystalline `rafts' with externally controllable properties
have been modeled with Monte-Carlo simulations, and the resulting
lattice constant, both experimental and simulated, has been compared
to the minimum in the effective interaction potential. In
treating the effective colloidal interactions, we find that a straightforward,
pairwise approach provides a reasonable description in describing
the lattice parameter. The minimum in the effective interaction formed
from combining the electrostatic repulsions and dipolar attractions
is a useful means to approximate the lattice parameter, although close
to melting fluctuations lead to a larger value in the lattice parameter
than this minimum. Conversely, at high field strengths, our treatment
appears to be less accurate. Apparently, higher order terms which
we have not accounted for may become important, such as non-linear
Poisson-Boltzmann contributions to the electrostatic interactions,
leading to a deviations from the Yukawa form {[}equation (\ref{eqYuk})]
\cite{brunner2002} or limitations in our pairwise treatment of the
dipolar attractions \cite{zahn2003}. Other possibilities include
inaccuracies in our assumptions for the effective colloid charge {[}equation
(\ref{eqZeff})], limitations of the 2D behaviour of the experimental
system. Perhaps due to a cancellation of errors, at high field strengths,
the minimum in the effective interaction can sometimes provide a more
accurate value of the lattice parameter than the simulations.

Another important assumption lies in our derivation on the electrostatic
interactions. While their Yukawa-like form is well-accepted \cite{alexander1984},
the values the effective colloid charge comes from equation (\ref{eqZeff}).
We note that the Debye length we determine seems rather constant across
the four particle sizes considered, {[}Fig.\ref{figKappaSigma} (a)],
which suggests that out approach is at least consistent. However challenging
it may be, a more quantitative measurement of the local ion concentration
in the vicinity of the crystalline rafts is desirable and will be
considered in the future. Furthermore, there may be some variation
in the value of $Z_{eff}$. In fact this should affect all state points
in a similar manner (for each particle size), and therefore we anticipate
a similar outcome, but that a different value might be arrived at
in the fitting of the Debye length, however, owing to the relatively
short-ranged nature of the interaction, the effect of a different
value for the effective charge is unlikely to impinge significantly
the Debye length and should have a negligible effect on our main results.

We have also assumed that a given system is described by
\emph{one single}Debye length. In fact, in the counterion-dominated
regime, the Debye length is in fact a function of colloid concentration
\cite{rojas2002}, an effect we have neglected. This might lead to
a tightening at higher colloid concentration (i.e. high field strength),
which may be expected to lead to an increase in deviation with the
experimental results (Figs. \ref{figSmallExpMC} and \ref{figBigExpMC}).
Nevertheless, varying the Debye length as a function of local colloid
concentration would be worth considering in the future.

We have extended our approach a range of colloid size, to optimise
the lattice tuneability for display applications. Tunable 2D crystal
rafts can behave as tunable diffraction gratings capable of filtering
white light into visible colours\citet{snoswell2006}. Similar tunable
diffraction has been proposed for display devices \cite{aschwanden2007}.The
range of colours obtainable for a given geometry is governed by the
lattice tunability. Fig 9b demonstrates that lattice tunability in
our system is maximised for particles of approximately 1 to 1.5 microns.
Only higher field strengths applied to smaller particles can increase
the lattice tunability. As already indicated, higher field strengths
are not practical in the current experimental system, however in future
experiments metal coated colloids would exhibit much stronger dipolar
interactions, enabling a reduction in field strength for a given attraction.
Regardless, our methodology illustrates using the well-developed machinery
of effective colloidal interactions, as a means to model potentially
complex interactions, useful for engineering purposes.

\newpage{} \bibliographystyle{naturemag}

\newpage


\vspace{3cm}

\noindent \textbf{Acknowledgments} We thank Jeremy Baumberg, Martin
Cryan, Daan Frenkel and Junpei Yamanaka for helpful discussions. CPR
acknowledges the Royal Society for funding and Hajime Tanaka for kind
provision of lab space and computer time. DES thanks Kodak European
Research Centre, Cambridge and EPSRC (EP/DO33047/1) for funding.)

\begin{table}
\begin{tabular}{cccccc}
\hline 
diameter &
Effective Yukawa &
$\kappa\sigma$ &
Ionic &
$\zeta$ potential &
melting well depth \tabularnewline
(nm) &
contact potential $\beta\epsilon$ &
&
strength (mMol) &
(mV) &
($k_{B}T$) \tabularnewline
\hline 
757$\pm$86 &
4.60$\pm$0.05$\times10^{3}$ &
19.0$\pm1.0$ &
0.06$\pm0.01$ &
-34.0 $\pm$ 1.4 &
2.7$\pm$0.1 \tabularnewline
965$\pm$44 &
5.78$\pm$0.02$\times10^{3}$ &
28.4$\pm2.0$ &
0.08$\pm0.01$ &
-37.1 $\pm$ 1.4 &
2.9$\pm$0.1 \tabularnewline
1390$\pm$30 &
8.27$\pm0.05\times10^{3}$ &
32.3$\pm3.0$ &
0.05$\pm0.005$ &
-39.9 $\pm$ 2.0 &
3.1$\pm$0.1 \tabularnewline
2070$\pm$40 &
1.17$\pm0.01\times10^{4}$ &
38.8$\pm3.0$ &
0.04$\pm0.005$ &
-43.6 $\pm$ 2.4 &
3.4$\pm$0.1\tabularnewline
\hline
\end{tabular}

\caption{Fitted parameters of the Debye screening length used in the MC simulations
for systems with different colloid diameters. Errors in the diameter
are standard deviations in the data obtained found from electron microscopy
measurements, while those in $\beta\epsilon$, $\kappa\sigma$, and
the ionic strength are estimated from the fitting of $d$ from MC
data.$\zeta$ potential measurements were made in a 0.01 mMol KCl
solution \label{tableKappaSigma} }
\end{table}

\newpage{}

\begin{figure}[htb]

\caption{The interaction potentials considered in this system. The Yukawa
repulsion $u_{yuk}(r)$ models the electrostatic repulsions {[}dotted
line, equation (\ref{eqYuk})], the dipolar attractions are plotted
as $u_{dip}(r)$ dashed line, {[}equation (\ref{eqDip})]. These are
combined to yield the total potential $u_{tot}(r)$, solid gray line
(equation \ref{eqU}).This figure corresponds to a particle diameter
of $\sigma=757$ nm and a field strength $E=30$ kVm$^{-1}$\label{figU}}
\end{figure}

\begin{figure}[htb]

\caption{A schematic of the experimental set-up, side view (a) and plan view
(b). The electric field is assumed to be constant across the observation
region, as shown in (b). (b) depicts the quadrupolar electrode configuration
(a,b),(c,d) \label{figSchematic}}
\end{figure}

\begin{figure}[htb]

\caption{Images of crystal rafts for the $\sigma=945$ nm system for applied
fields of 29.0 and 80.0 $kVm^{-1}$for \textbf{a} and \textbf{b} respectively.
Scale bars=10 $\mu m$.\textbf{c} snapshot from MC simulation for
same system for a field of 30.0 $kVm^{-1}$and $N$=144. \label{figPix}}
\end{figure}

\begin{figure}[htb]

\caption{Change in the lattice constant for simulated systems of different
sizes. Here $\sigma=945$ $nm$ and a field strength $E=20$ $kVm^{-1}$.
Each run was repeated four times except $N=576$. Error bars denote
standard deviations over the results from different runs. Dashed line
is solely a guide to the eye.\label{figFiniteSize}}
\end{figure}

\begin{figure}[htb]

\caption{(a) The Debye length $\kappa^{-1}$ fitted from Monte-Carlo simulation
as a function of particle size. Dotted line denotes $\kappa^{-1}$=40
nm taken for the simulations in section \ref{tunability}. Error bars
denote the uncertainty in the Debye length resulting from the fitting
in Figs. \ref{figSmallExpMC} and \ref{figBigExpMC}. (b) The contact
potential of the Yukawa interaction $\beta\varepsilon_{yuk}$ {[}equation
(\ref{eqEpsilonYuk})] as a function of particle size. We include
$\sigma=400$ nm for which we take the Debye length $\kappa^{-1}=40$
$nm$. }

The straight line is fit of the function $\beta\varepsilon_{yuk}(\sigma)=A\sigma^{b}$
where the fitted parameters are $A=11.1\pm1.1$ and $b=0.91\pm0.01$.
\label{figKappaSigma} 
\end{figure}

\begin{figure}[htb]

\caption{Radial distribution functions for an applied field of $E$=30.7 $kVm^{-1}$
for the $\sigma=2070$ nm system. Circles are experimental data, line
is MC simulation for $N$=144. Arrows denote the first few peaks of
a hexagonal lattice, $d,$$\sqrt{3}d,2d,\sqrt{7}d,3d$...}

\label{figG} 
\end{figure}

\begin{figure}[htb]

\caption{Experimental results compared to MC simulation and the minimum in
the effective potential. \textbf{a} 757 nm and \textbf{b} 945 nm diameter
colloids. Filled squares are $N=72$ while crossed are $N=144$ from
MC simulation. Circles are experimental data. Grey lines are the minimum
of equation \ref{eqU}. Error bars in the experimental data are standard
deviations. \label{figSmallExpMC}}
\end{figure}

\begin{figure}[htb]

\caption{Experimental results compared to MC simulation and the minimum in
the effective potential. \textbf{a} 1390 nm and \textbf{b} 2070 nm
diameter colloids. Filled squares are $N=72$ while crossed are $N=144$
from MC simulation. Circles are experimental data. Grey lines are
the minimum of equation \ref{eqU}. Error bars in the experimental
data are standard deviations. \label{figBigExpMC}}
\end{figure}

\begin{figure}[htb]

\caption{(a) The lattice spacing around melting ($dm$, filled squares), and
at at applied field of 100 $kVm^{-1}$, as determined from MC simulation
(crosses) and from Equation \ref{eqU}, as a function of particle
size. Inset shows a radial distribution function close to melting
($\sigma=1000$ nm, $E=11$.0 $kVm^{-1}$). Arrow indicates split
second peak.}

(b) the tunability ratio as a function of particle size. Connecting
lines are guides to the eye.\label{figTuning} 
\end{figure}

\newpage{}

\begin{figure}[htb]
 \includegraphics{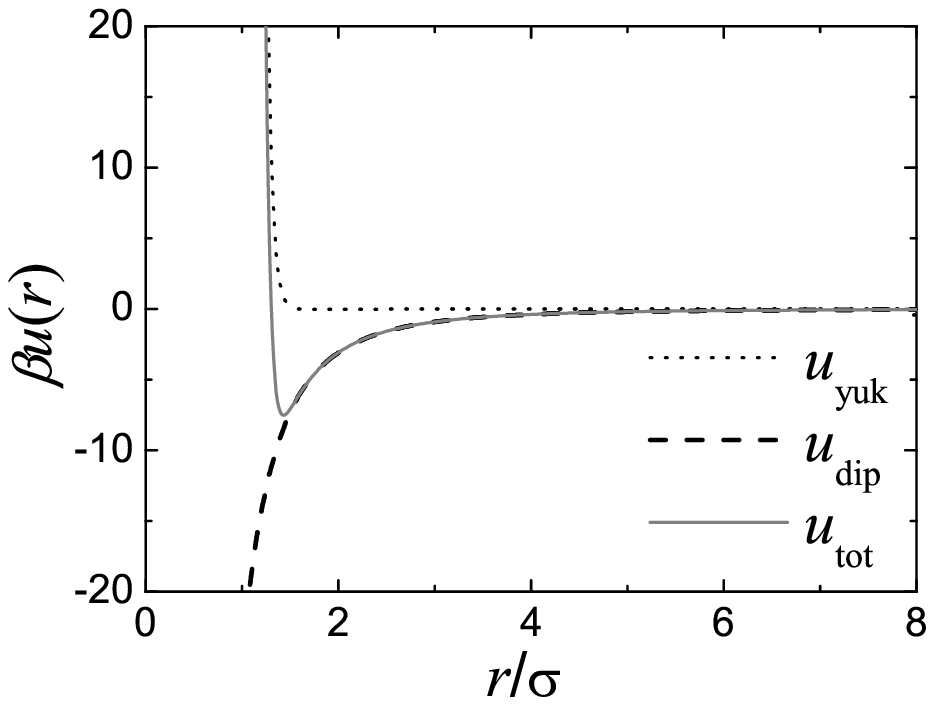}
\end{figure}

\begin{figure}[htb]
 \includegraphics{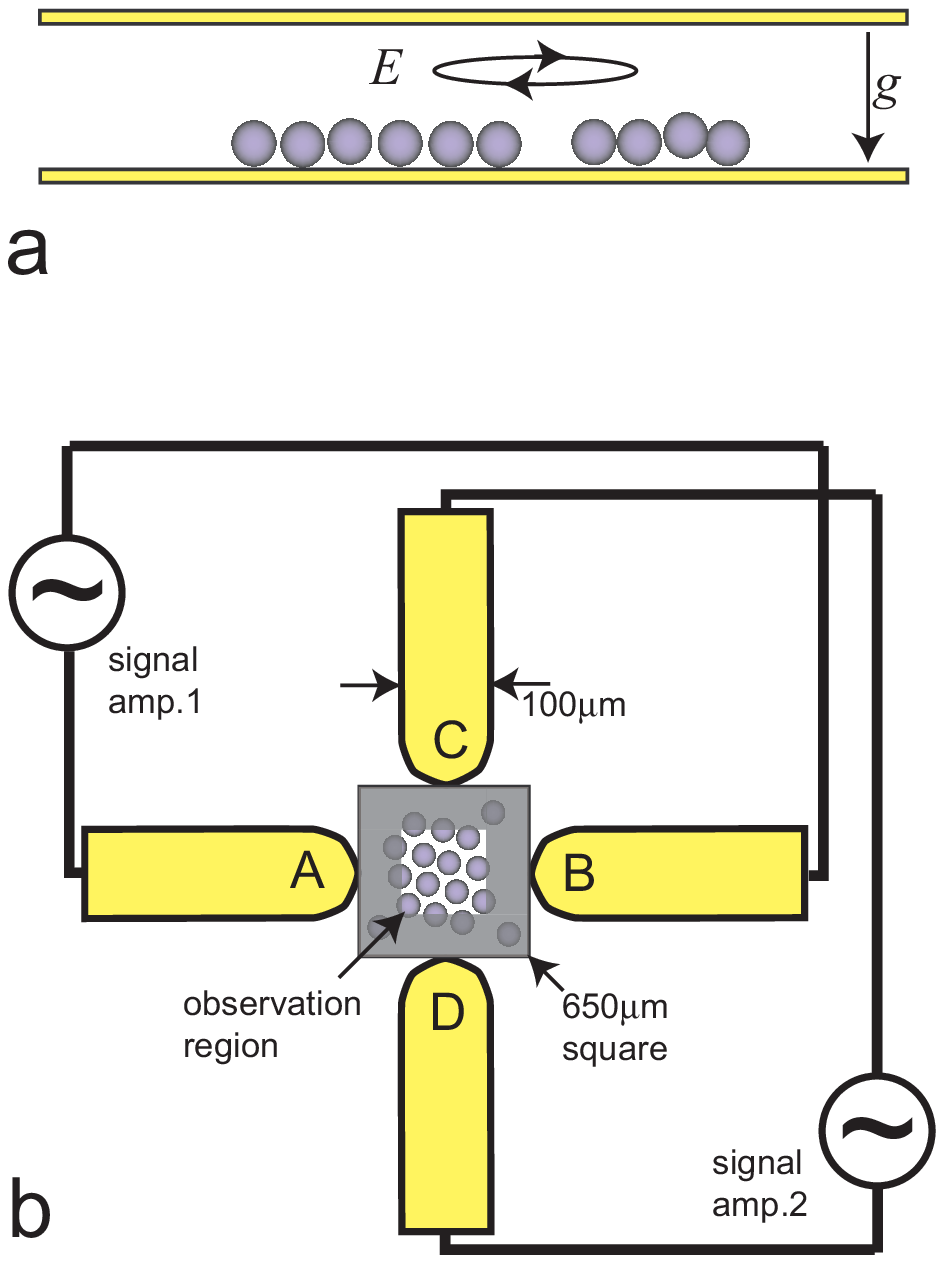}
\end{figure}

\begin{figure}[htb]
 \includegraphics{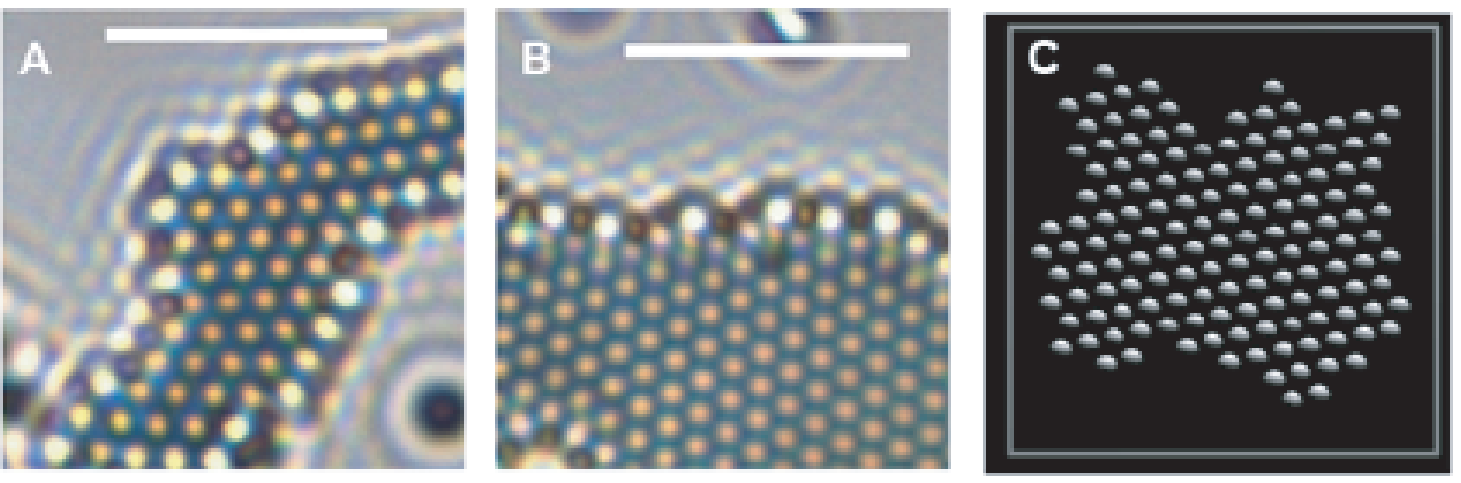}
\end{figure}

\begin{figure}[htb]
 \includegraphics{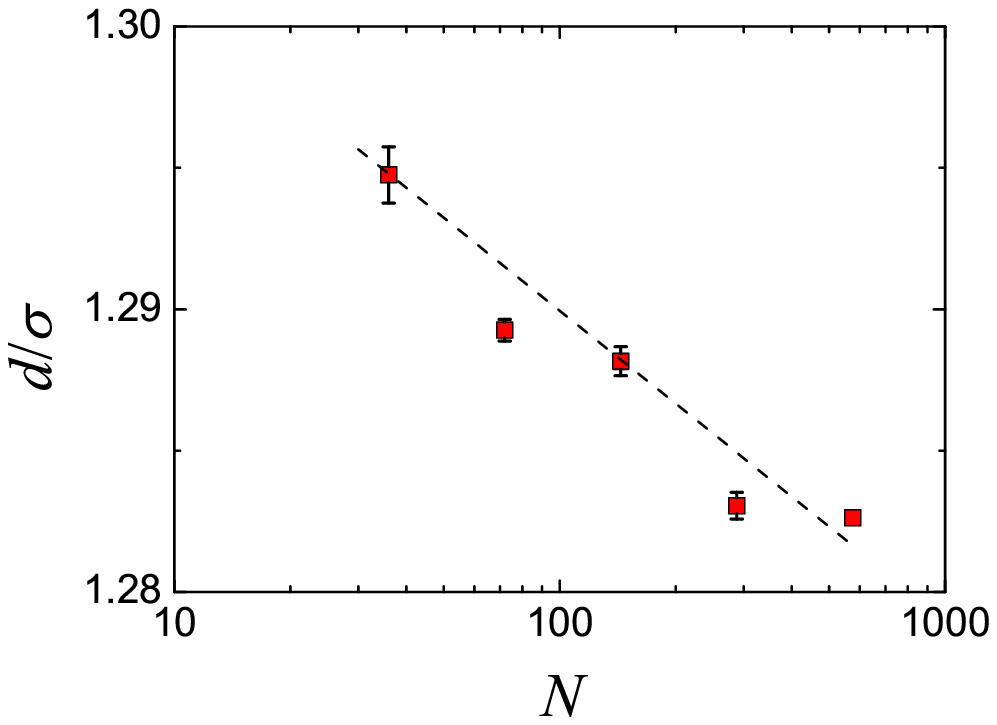}
\end{figure}

\begin{figure}[htb]
 \includegraphics[scale=0.7]{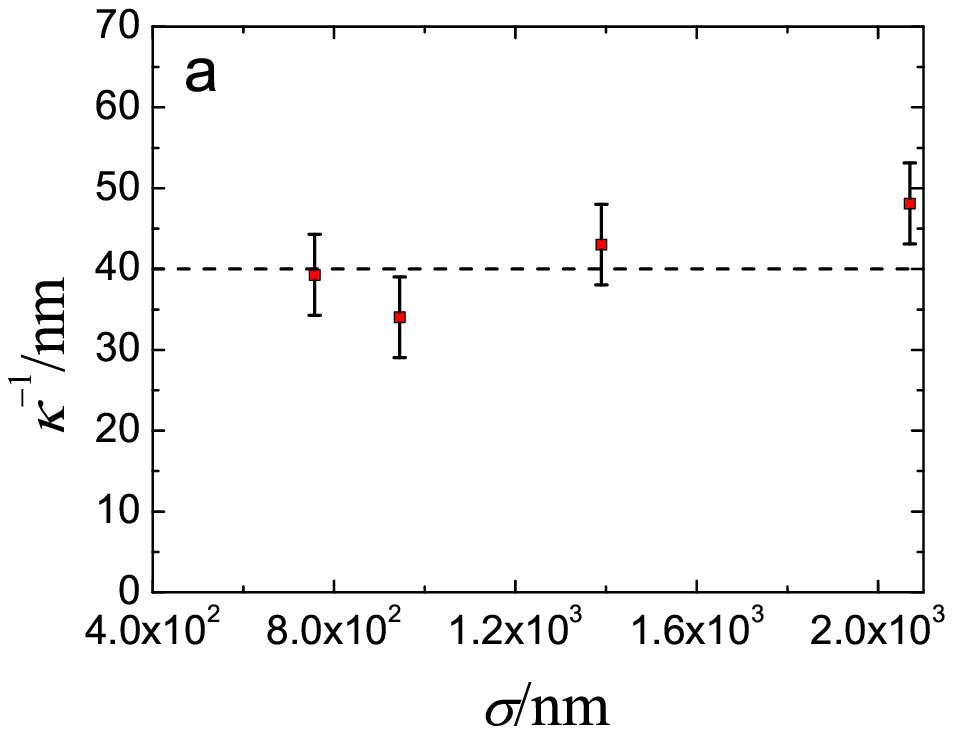}\includegraphics[scale=0.7]{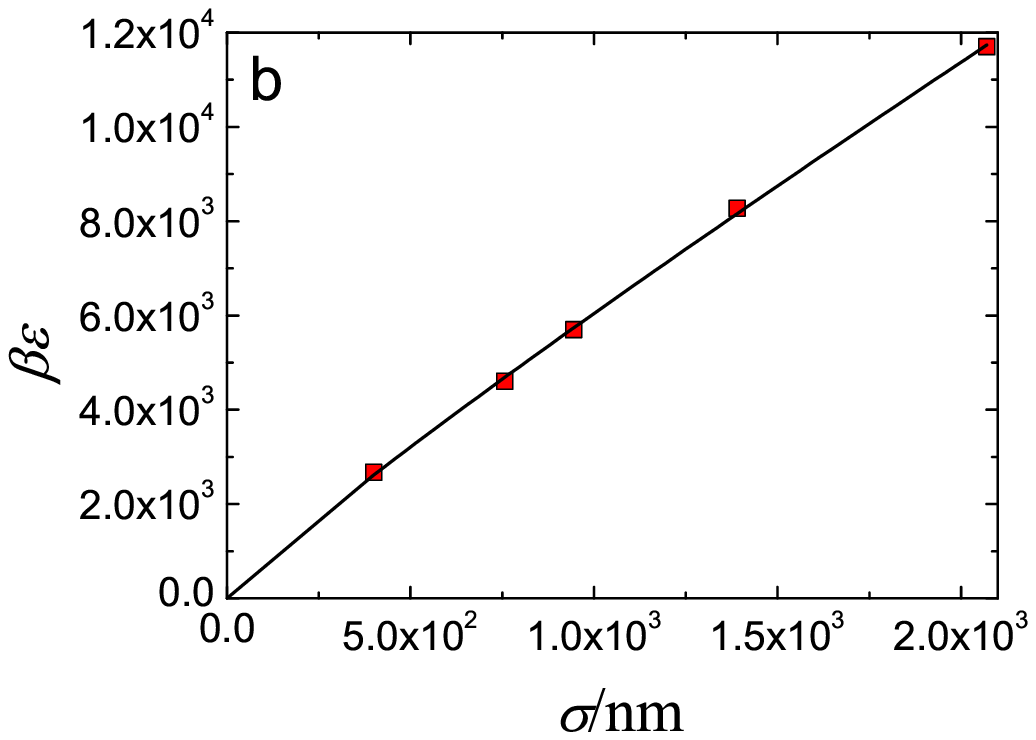}
\end{figure}

\begin{figure}[htb]
 \includegraphics{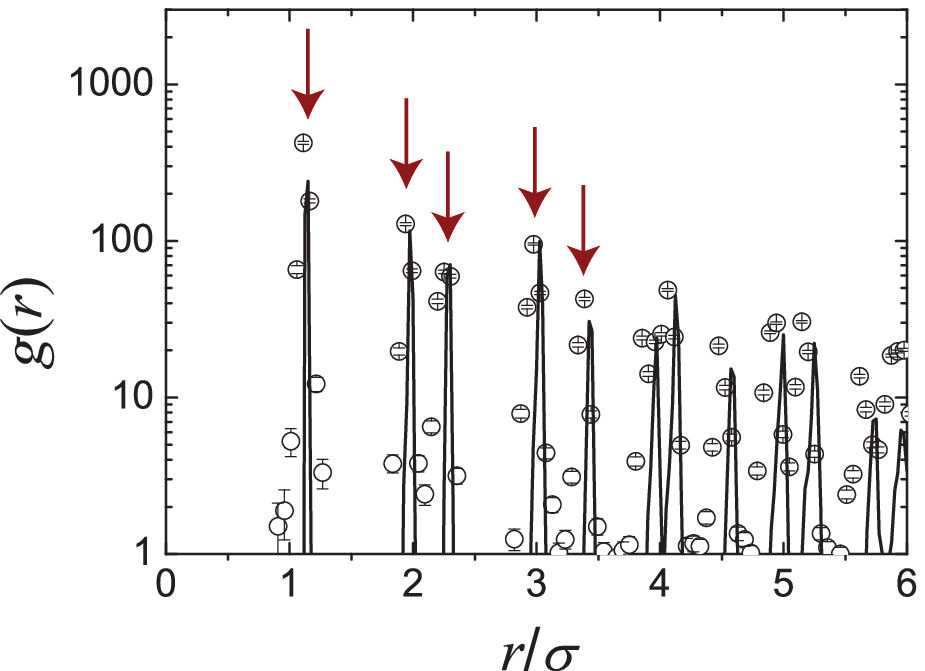}
\end{figure}

\begin{figure}[htb]
  \includegraphics[scale=0.7]{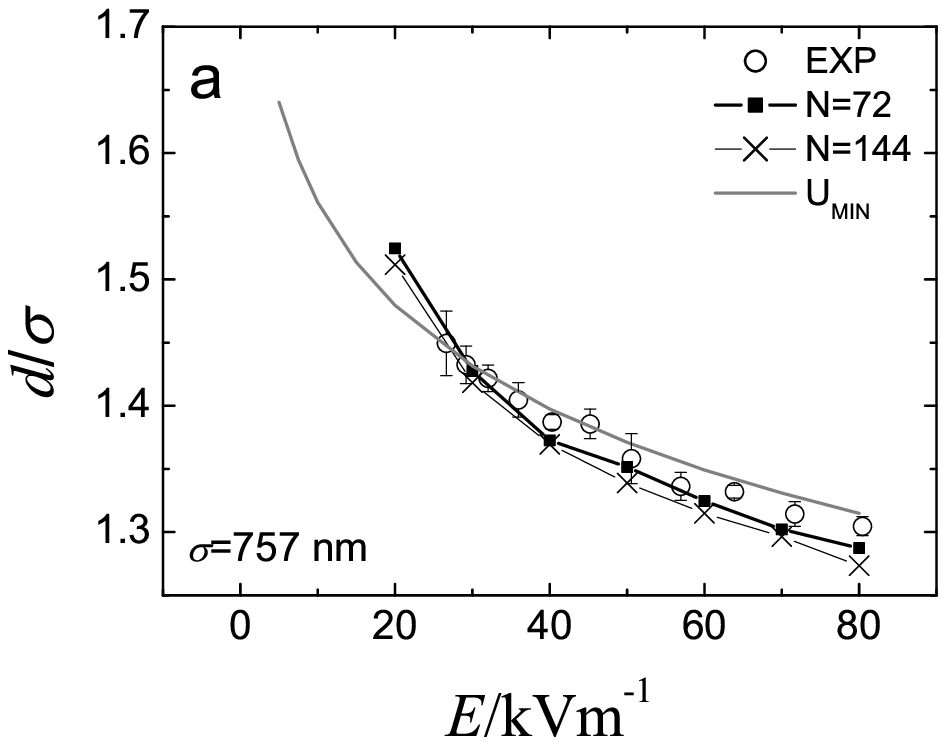}\includegraphics[scale=0.7]{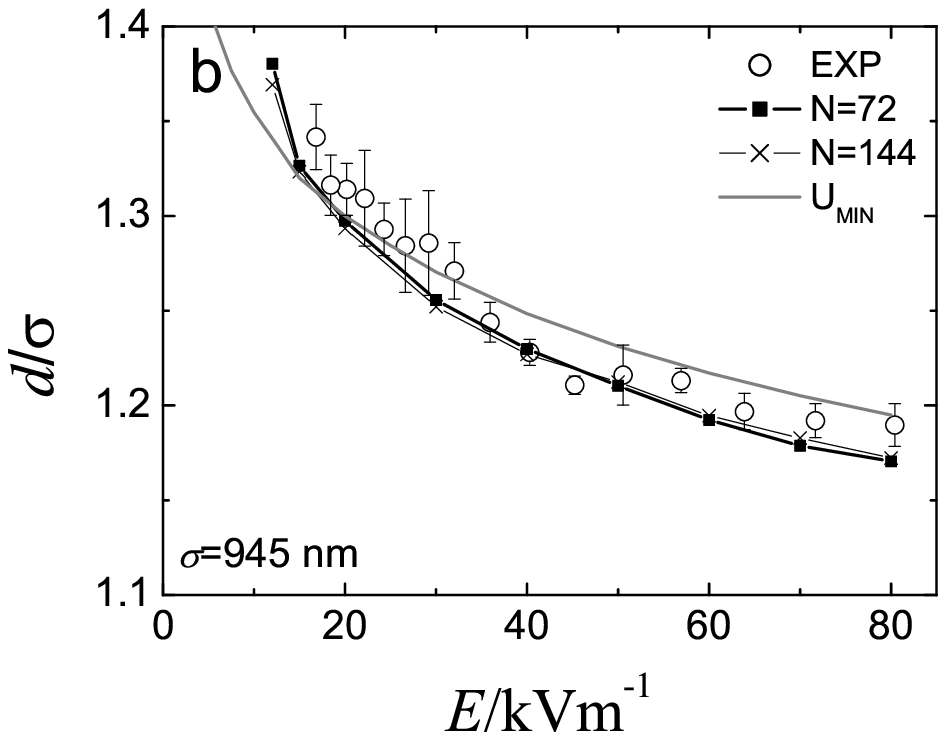}
\end{figure}

\begin{figure}[htb]
 \includegraphics[scale=0.7]{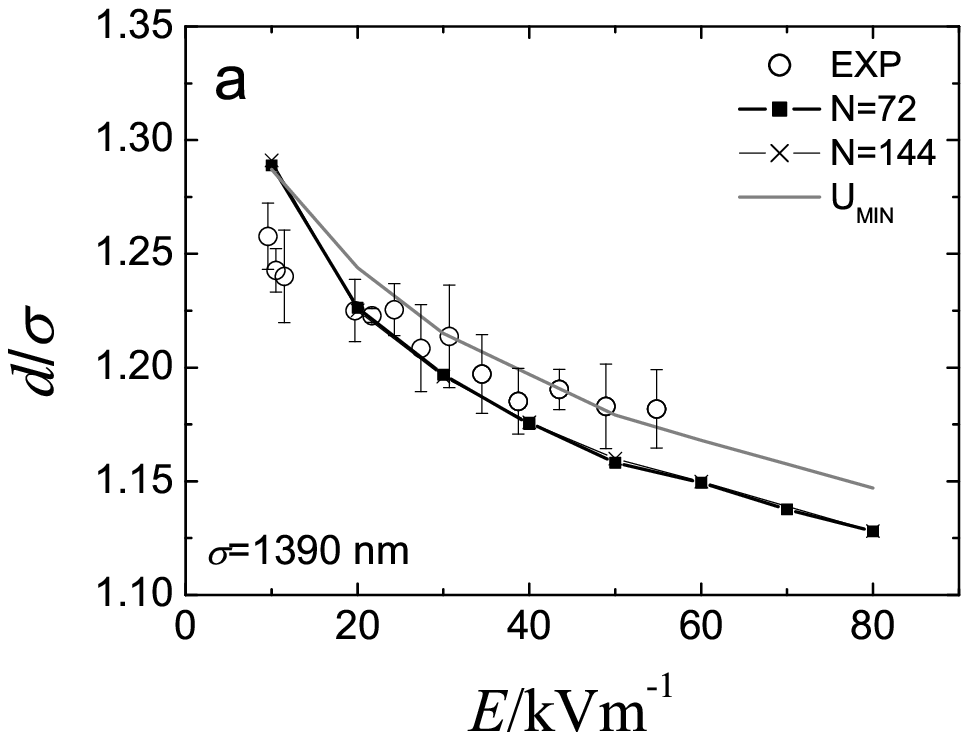}\includegraphics[scale=0.7]{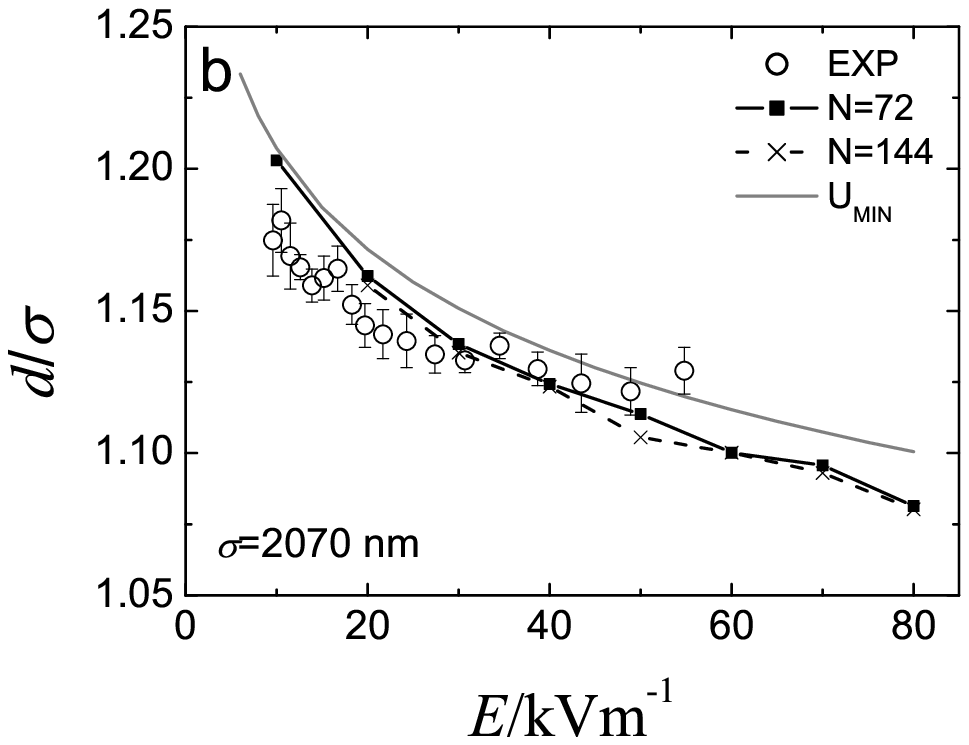}
\end{figure}

\begin{figure}[htb]
 \includegraphics[scale=0.7]{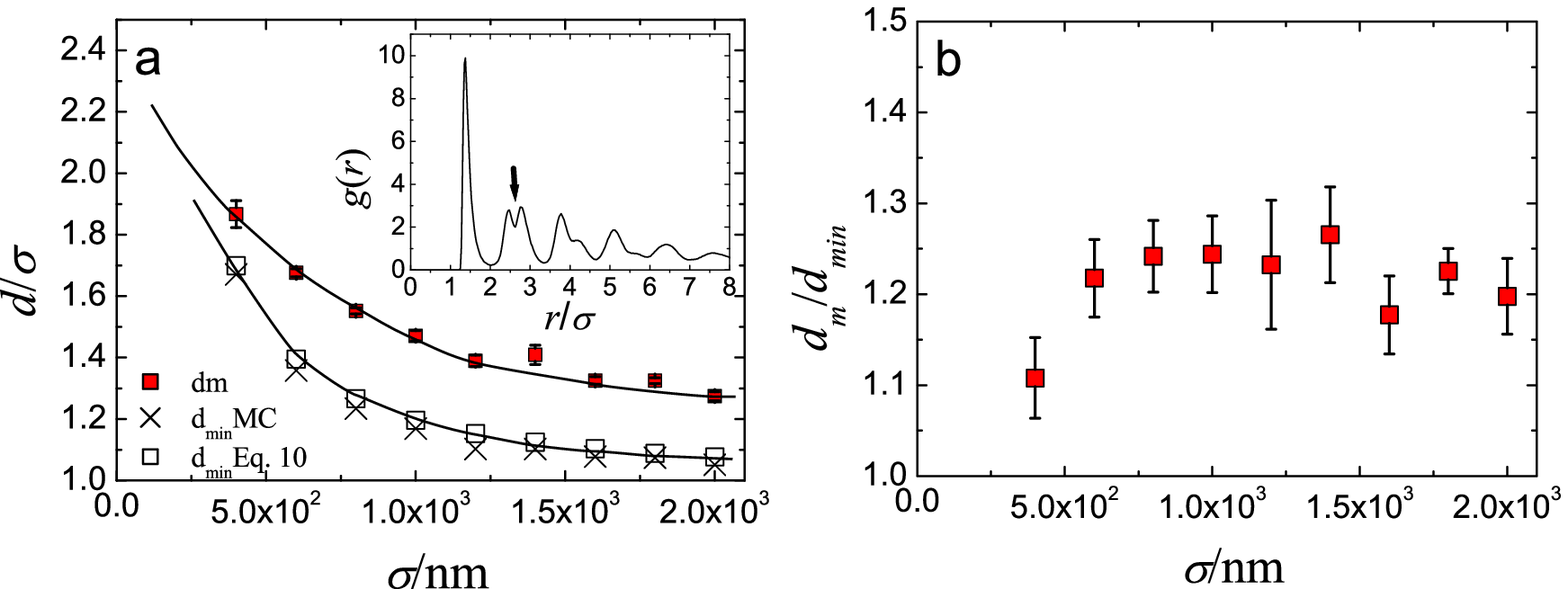} 
\end{figure}

\end{document}